\documentclass{cimento}
\usepackage{graphicx}        
\usepackage{color}           
\usepackage{url}             

\title{Airborne observation of 2011 Draconids meteor outburst: the Italian mission}

\author{C.~Sigismondi\from{ins:i}\from{ins:a}\from{ins:u}\from{ins:e}\from{ins:f}}

   \instlist{\inst{ins:i} ICRA, International Center for Relativistic Astrophysics, and Sapienza University of Rome, P.le Aldo Moro 5 00185, Roma (Italy)
                    
             \inst{ins:a} Universit\'e de Nice-Sophia Antipolis (France)
                     
             \inst{ins:u} Istituto Ricerche Solari di Locarno (Switzerland)
                    
             \inst{ins:e} GPA, Observatorio Nacional, Rio  de Janeiro (Brasil)
                   
             \inst{ins:f} IMO, International Meteor Organization 
             }
\PACSes{\PACSit{96.25.-f}{Planetology of comets and small bodies}
\PACSit{96.30.Za}{Meteors, meteoroids, and meteor streams}}

\begin{document}

\maketitle

\begin{abstract}

The outburst of 8 October 2011 of Draconids meteors has been observed visually onboard of Alitalia AZ790 flight. The enhanced zenithal hourly rate around ZHR=300 from 19 UT to 21:50 UT has been observed over central Asia. The data and the method of analysis are described and compared with other observations made worldwide.
 
\end{abstract}

\section{Introduction}
Meteor showers have been observed and recorded since the time of early Chinese astronomers.\cite{Jenniskens,Maffei,Yoke}
Giovanni Virginio Schiaparelli, Italian astronomer in Milan during the nineteenth century, discovered the link between comets and meteor showers.\cite{Flammarion} 
The parent comet of Draconids meteor shower is the periodical comet P/21 Giacobini-Zinner; and the passages of 1900 with minor contribution of 1887 and 1894 passages\cite{Maslov} delivered the meteoroids which have produced the outburst of october 8, 2011.

In the last decades the interest around these phenomena increased after the extraordinary fireballs shower of Leonids in 1998,\cite{Sigismondi98} which lead to study the dynamics of meteors trails left after the past passages at perihelion of the parent comet. This study, in the case of Leonids, involved passages back to the thirteenth century. 
Meteor showers are a significant part of space weather arguments, being important their prediction for protecting the satellites by orienting their instruments in the opposite direction of the radiant for the duration of the phenomenon.
Also the field of research on lunar impacts has been opened recently, because of the increased probability to observe such an impact from the Earth during meteor showers.\cite{Sigismondi2000a,Sigismondi2000b} Several events have been recorded in order to better understand the physics of hypervelocity impacts. 

\section{The 2011 Draconids outburst}

The forecast of a remarkable outburst of Draconids meteors, associated to P/21 Giacobini-Zinner comet, for 8 october 2011, has been included in the research programs of ICRANet by fixing the calendar of its international acitivities in China to fit also this astronomical event. The flight Alitalia AZ790 from Rome to Beijing, to attend the Third Galileo-Xu meeting\cite{ICRANet} between chinese and italian experts on relativistic, stellar and solar astrophysics was perfect for this purpose.
This flight going over northern Asia during the night has been exploited for a visual observational session during the 7 hours around the predicted peaks.
Althougth the field of view of the observations and the lenght of the astronomical night were reduced with respect to normal Earth-based observations, the data gathered were statistically significant to monitor the phenomenon.
In the spirit of the science meeting, this observation has been a tribute to the ancient chinese astronomers: thanks to their reports several meteor showers outburst have been studied.
A data report is presented and compared with other international missions.

\medskip
A zenithal hourly rate of $\sim7000$ has been predicted for the Draconids outburst of October 8, 2011,\cite{IMCCE} collocating this phenomenon at the level of the most abundant showers ever observed.
French CNRS organized a flight over the North Polar regions, with a team of international scientists leading worldwide the field of research on minor bodies of solar system.\cite{IMCCE,SETI}

\medskip
ICRANet promoted this research by fixing my flight AZ790 to Beijing, in the occasion of the third Galileo-Xu meeting, on the night of the event, because since 1998 I am a contributing observer to IMO, International Meteor Organization, with SIGCO code. I developped an expertise on meteors observations during major and minor meteor showers along these years. 

\section{Observing method}

The well known restrictions in an airline flight, in terms of space and field of view available from the windows, suggest to limitate the observation to the naked eye.
The double windows of the airplane enhance the multiple reflections when a videocamera is not perpendicular to its semi-transparent surfaces. Moreover the presence of the full Moon in the sky gives another source of light beyond the internal light of the airplane.
In order to avoid Moonlight, the seat A24 with the window facing North has been selected.
Scattered Moonlight was present up to 20:40 UT. After this time the Moon set and the sky was perfectly clear, and the Gerbert's star\cite{gerbertus} (HR4893) of magnitude $m_v$=5.35 was visible. Before the faintest star visible was 4 Ursae Minoris of $m_v$=4.8. The limiting magnitude increased of 0.5 magnitudes has been included in the analysis of fig. 2. Before 20:40 UT all data in the lower luminosity class have been multiplied by 1.5, the square root of 2.2, the observed population index of these Draconids.  

In order to reduce the light from the cabin, reflected by the glasses of the window, the green Alitalia blanket has been fixed on the frame of the window in order to create a black box with my head inside; the flux of the air conditioned has been sent into this box in order to prevent the creation of dew on the glass. The distance from the glass and the eye has been measured in order to evaluate the available field of view.
The field of view has been measured also by calculating the angular distance (with the Stellarium 10.3\cite{Stellarium} goniometer) between the stars visible at its borders and noted for a refined analysis. 

The initial prediction of $\sim 7000$ meteor per hour would suggest in any case to observe within a limited field of view in order to be able to record the phenomen by naked eye.
Uninterrupted observing sessions of 10 minutes have been realized, as in the case of 1998 Leonids, when I made 15 minutes session. About ten seconds for changing position and writing the data was used between one session to the following. Typically no more than 7 meteors were observed during each session. 
The number, the direction and the luminosity of the meteors observed and memorized have been noted, from the beginning of twilight between dusk and dawn, for a total of more than 7 hours of continuous observation.
An audio recording system was ready to record the phenomenon in case of exceptionally high rates of meteors, but this was not necessary. 
I noticed also the presence of several Orionids meteors all night long: 20 Orionids recorded, much faster, as well as 4 ohter meteors: 2 true sporadic and 2 belonging to another active shower, probably the delta Aurigids or the October Arietids peaking also on 8 October.  
It is possible that Orionids have tracks systematically longer than Draconids in my restricted field of view and this fact rose their probability of appearance.
\begin{figure}
\centerline{\includegraphics[width=1\textwidth,clip=]{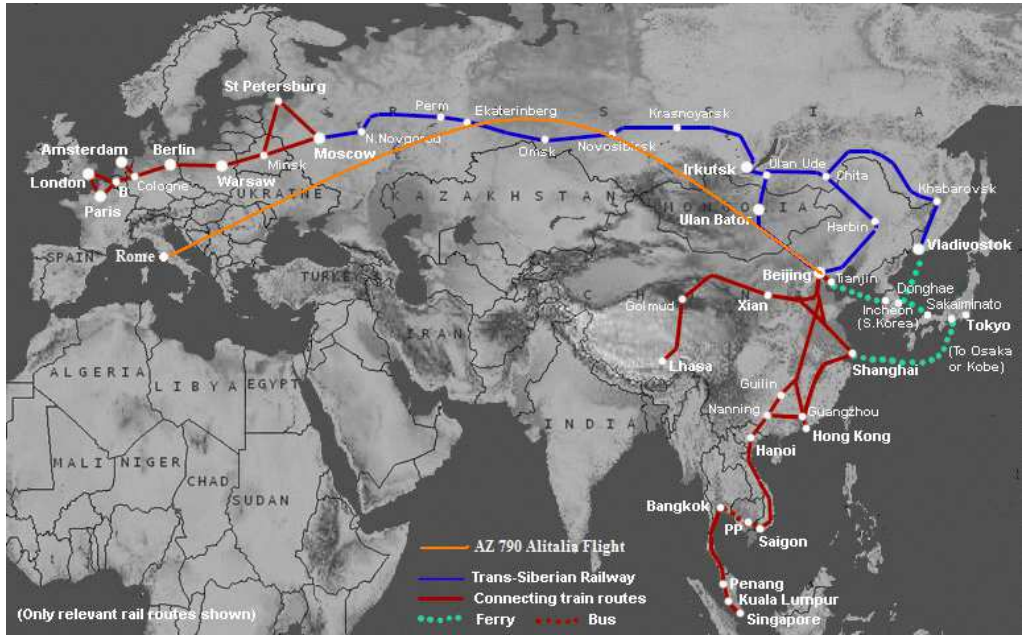}}
\caption{The AZ790 flight followed the trans-Siberian railway path and trans-Mongolian path after Ekaterinburg. The sunset occurred at 15:46 UT over Brasov in Romania, and the sunrise at 22:39 UT over Mongolia.}
\label{Fig. 1}
\end{figure}

\begin{figure}
\centerline{\includegraphics[width=1\textwidth,clip=]{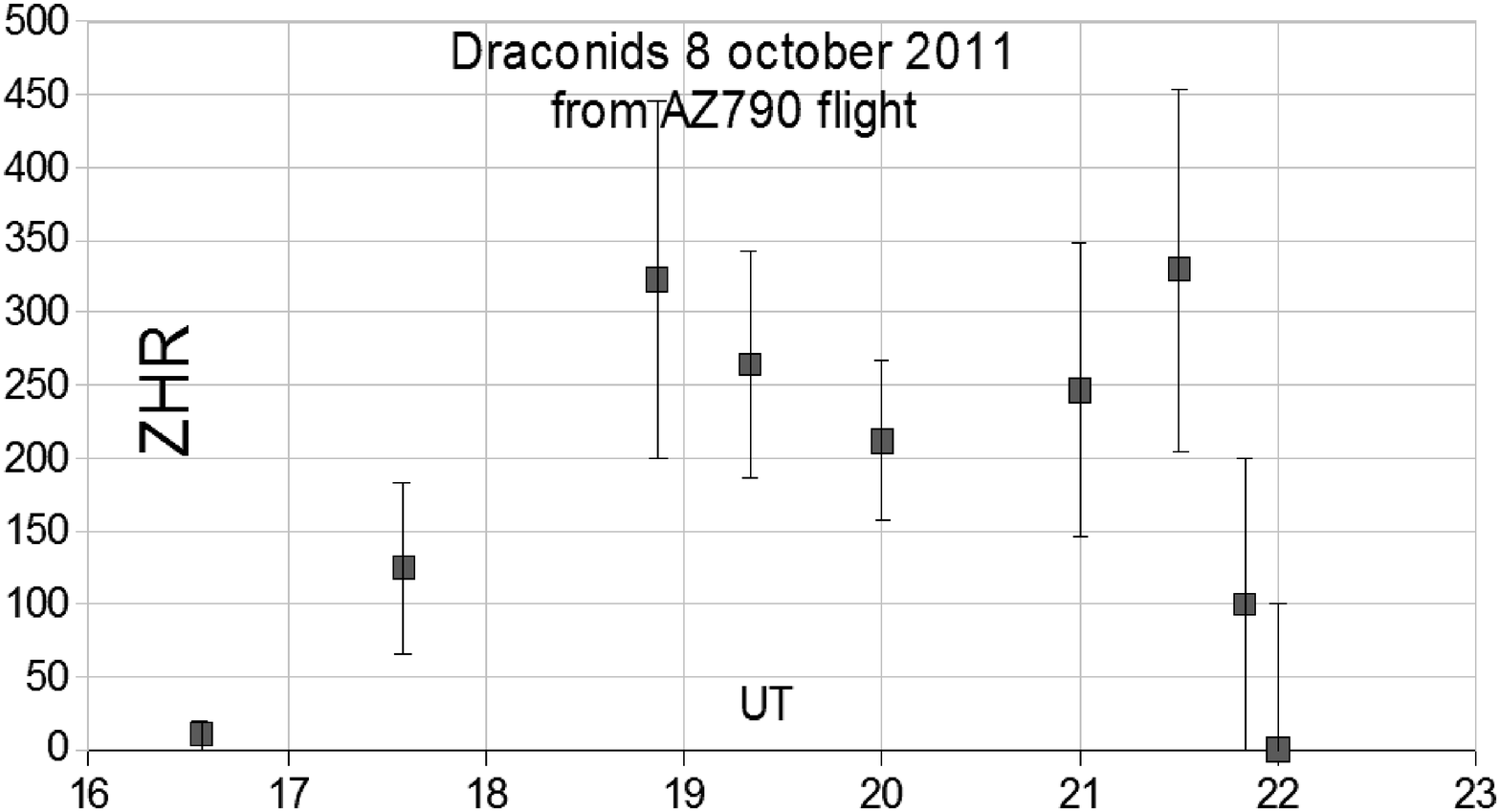}}
\caption{The Zenithal hourly rate of Draconids shower recorded during the Alitalia AZ790 flight of october 8-9, 2011. The errorbars have been calculated with Poisson's statistics.}
\label{Fig. 2}
\end{figure}

\section{Data Reduction and Conclusions}

The trajectory of the airplane has been divided into time channels according to the notes taken onboard. For each channel the central latitude has been considered in order to calculate the height of the radiant above the horizon. The duration of the night perceived onboard the Airbus 330 has been of 6h 53m from sunset to sunrise; the corresponding duration for the reference location ($55^o$N, $65^o$E) where we flew at local midnight was 12h 50m. I computed the height of the radiant above the horizon with Ephemvga\cite{Ephemvga} at that reference location using the local time and correcting by the changing latitude of the airplane. The ZHR has been calculated by using the observed times, while the height of the radiant has been computed with a proportional time, i. e. 1 hour after the airplane midnight corresponds to about 2 hours of celestial sphere rotation, according to the ratio 12h50m/6h52m. All the measurements have been translated into UT timings.
The US standard atmosphere\cite{US1976} has been used for calculating the refraction of the residual atmosphere. 
The Zenithal Hourly Rate, ZHR, of Draconids on October 8, 2011 was lower than the predictions, peaking at around 300 from 19 to 21:30 UT with two peaks of activity detected.
For calculating the population index r, the ratio of the number of meteors in subsequent magnitude channels, the number of meteors recorded was too low, nevertheless the population index has been evaluated as $r=2.2\pm0.7$ by using the first four luminosity classes, and eliminating the last one.\cite{erre} The last luminosity class was clearly uncomplete.
To obtain the final ZHR, the number of meteors in the last luminosity class and observed before 21:40 UT have been multiplied by 1.5=$\sqrt r$ in order to take into account the higher limiting magnitude ($\ge0.5$ mag. i. e. half luminosity class) experienced because of Moonlight. 

\begin{table}
  \caption{Population index of 47 draconids}

  \begin{tabular}{ccc}
    \hline
      magnitude     & counts   & population index    \\
    \hline
      -2  &  1   &       /   \\
      -1  &  2   &      2  \\
       0  &  5   &      2.5    \\
       1  &  15  &      3    \\
       2  &  19  &      1.27    \\
       3  &  5   &       /    \\
         &  tot=47   &     average r=2.2$\pm$0.7    \\
    \hline
  \end{tabular}
\end{table}

The first meteor has been observed at 16:34 UT, it was $Mv=-2$. It could belong to the 1887 trail, according to the predictions of Maslov\cite{Maslov}. Afterwards the number of events remained low until 18:35 UT when they started to increase.    

The velocity of the airplane of $v\sim 0.3 $km/s eastward would have produced a shift of $1^o$ eastward in the position of the radiant,
because of the aberration-like phenomenon of the vectorial composition with the velocities of the meteors v=20.9 km/s. 
The radiant of the Draconids was in my field of view, the visual estimate of its position was within $10^o$ from its calculated position, but the absence of photo prevented a better accuracy and the verification of the kinematical aberration of the radiant.
  
\medskip
The results of visual observation conducted onboard of Alitalia flight AZ790 has been confirmed by other reports worldwide.\cite{IMCCE,SETI} 
From visual observations of 2164 Draconids collected by the International Meteor
Organization, G. Barentsen calculated a peak Zenith Hourly Rate (ZHR) of $300\pm 30$ Draconids per hour at Oct. 8 $20:03\pm 10$min UT with a shower full-width-at-half-maximum (FWHM) of 100 minutes.  The shower was rich in faint meteors,
with a population index of r=$2.80 \pm 0.05$.\cite{Green} 
The double peak structure visible in fig. 2 can be a local situation within the comet's trail of meteoroids.
A similar occurrence was observed during the 1998 Leonids.\cite{Sigismondi98} The sum of the counts over several observers showed that the two peaks of 1998 Leonids reflected a local structure of the trail, subjected to the attraction of the Earth, during past close encounters.
The FWHM observed by IMO is also very well confirmed by our airborne observations.



\bibliography{draconid}
   
\bibliographystyle{varenna}

\end{document}